\documentclass[%
reprint,
 amsmath,amssymb,
 aps, prl,
]{revtex4-2}

\usepackage{graphicx}
\usepackage{dcolumn}
\usepackage{bm}

\usepackage{color}
\usepackage{comment}
\usepackage{here}

\newcommand{\bra}{\left\langle}
\newcommand{\ket}{\right\rangle}

\newcommand{\bv}[1]{{\boldsymbol #1}}
\newcommand{\ep}{\epsilon}

\renewcommand{\ss}{{\rm ss}}

\begin{document}


\title{Microscopic cut-off dependence of an entropic force in
  interface propagation of stochastic order parameter dynamics}


\author{Yutaro Kado and Shin-ichi Sasa}
\affiliation{Department of Physics, Kyoto University, Kyoto 606-8502, Japan }
\date{\today}

\begin{abstract}
The steady propagation of a $(d-1)$-dimensional planer interface in $d$-dimensional space is studied by analyzing mesoscopic non-conserved order parameter
dynamics with two local minima under the influence of thermal noise.
In this analysis, 
an entropic force generating interface propagation is formulated using a perturbation method. It is found that the entropic force singularly
depends on an ultraviolet cut-off when $d \ge 2$. The theoretical calculation
is confirmed by numerical simulations with $d=2$. The result
means that an experimental measurement of the entropic force provides
an estimation of the microscopic cut-off of the mesoscopic description.
\end{abstract}

\maketitle


{\it Introduction.---}
Macroscopic dynamics in nature are often described by deterministic equations. If a class of
phenomena is found to be described by a simple deterministic equation with a few parameters,
the phenomena can be studied by analyzing the universal equation. Fluid dynamics is a typical example of the success of such an approach \cite{Fluid-dynamics}. However, there are cases in which simple deterministic equations cannot be found. For example, the macroscopic motion of locally conserved quantities in one or two dimensions cannot be described by a hydrodynamic equation with finite transportation coefficients \cite{FNS}. Another example is the dynamical behavior near a critical point, where transport coefficients show a singular behavior \cite{Kadanoff1968, Kawasaki1970}.

Even in such cases, a mesoscopic model with thermal noise can describe the macroscopic dynamical behavior. For example, fluctuating hydrodynamics can correctly describe the singular behavior of hydrodynamics in low dimensions \cite{FNS}. The dynamical behavior near a critical point also can be described by the Ginzburg-Landau model with thermal noise \cite{Hohenberg_Halperin}. Here, fluctuations modify the mean-field properties given by the mesoscopic free energy, which includes the transition type \cite{Bak1976} as well as the critical exponents \cite{Hohenberg_Halperin, Wilson1974}. These examples show that fluctuation effects at mesoscopic scales lead to the renormalization of model parameters and that the lack of a simple macroscopic equation is connected to the infrared divergence of renormalized parameters. In such a case, phenomena observed at macroscopic scales cannot be separated from those at mesoscopic scales. 

The purpose of this Letter is to report another violation of this scale separation, which is qualitatively different from the previously known cases.  We study a steady propagation of a $(d-1)$-dimensional planer interface separating two phases in $d$-dimensions by analyzing the mesoscopic non-conserved order parameter dynamics with weak thermal noise. One remarkable finding is that the propagation velocity depends on the noise intensity, as already reported in the case $d=1$ \cite{Costantini2001}. This means that the noise modifies the average dynamic behavior. Although this is an interesting phenomenon, the mechanism behind it is simple. The driving force of the interface motion is the free energy difference between the two phases, which contains the entropic contribution in addition to the mesoscopic free energy given by the mesoscopic model. Here, the entropic contribution is expressed by fluctuation properties coming from the noise. Thus, if there is no special symmetry between the two phases, the propagation velocity depends on the noise intensity. In a special case where the mesoscopic free energy takes the same value in the two phases, the propagation occurs only as a result of the entropic force. The behavior can be understood by the renormalization of the mesoscopic free energy. The main message of this Letter is that the entropic force diverges when an ultra-violet cut-off goes to infinity for $ d \ge 2$. 

This result means that the propagation velocity driven by the entropic force depends on the microscopic cut-off. In other words, the mesoscopic description cannot be separated from a more microscopic system. We demonstrate this result using a theoretical calculation of the entropic force. Furthermore, we confirm this claim by performing numerical simulations of the order parameter dynamics with noise. We expect that this singular behavior is also observed in experiments such as interface motion in a spin-crossover complex \cite{Boukheddaden2016}. Surprisingly, an experimental measurement of the entropic force provides an estimation of the microscopic cut-off of the mesoscopic description.

{\it Setup.---}
For simplicity, we present the system in two dimensions. The generalization to
the other dimensions is straightforward. Let $\bm{r}=(x,y) \in \mathbb{R}^2$
be a position in a two-dimensional region $D \equiv [-L,L] \times
[0, L_y]$. $L$ and $L_y$ are sufficiently large and $L$ is assumed to
be infinity in theoretical arguments. 
We define a real scalar order parameter field $\phi(\bm{r},t)$
in the region $D$. The free energy functional of $\phi$ is given by 
\begin{equation}
  {\cal F}(\phi)=\int_{D} d^2\bm{r} \left\{f(\phi)
+ \frac{\kappa}{2}[(\partial_x \phi)^2 + (\partial_y \phi)^2] \right\} ,
\end{equation}
where $f(\phi)$ is a mesoscopic free energy density and $\kappa$ is a constant
characterizing the interface energy. 
Following the Onsager Principle, we assume that the dynamics of
$\phi(\bm{r},t)$ is described by
\begin{equation}
\partial_{t}\phi=-\Gamma \frac{\delta {\cal F}}{\delta \phi}
  + \eta,
\label{model}
\end{equation}
where $\Gamma$ is a constant representing the mobility and
$\eta$ is Gaussian white noise satisfying $\langle \eta(\bm{r},t)\rangle =0$
and the fluctuation-dissipation relation
\begin{equation}
  \langle \eta(\bm{r},t) \eta(\bm{r'},t')
  \rangle = 2\Gamma T \delta(\bm{r} - \bm{r'})\delta(t-t').
\end{equation}
This model has been referred to as "Model A" \cite{Hohenberg_Halperin}, which describes the non-conserved order parameter dynamics. 
More precisely, because this model describes mesoscopic dynamics, we
introduce a microscopic cut-off length  $2\pi/k_c$, where the amplitude
of the Fourier mode with $|\bv{k}| > k_c$ is set to zero.  


Specifically, we study a system in which the mesoscopic free energy density
$f(\phi)$ has two local minima at $\phi_1$ and $\phi_2$. We assume
$\phi_1 < \phi_2$ without loss of generality. 
We impose periodic boundary conditions in the $y$-direction and
$\phi(\bv{r})=\phi_1$ at $x=-L$ and $\phi(\bv{r})=\phi_2$
at $x=L$. A one-dimensional planer interface is initially
prepared at $x=0$. We then observe the motion of the interface.
The goal here is to determine the expectation value of the
steady propagating velocity of the interface. 

We first consider the case $T=0$ with $f(\phi)$ fixed. 
The model given by (\ref{model}) becomes a deterministic equation. The steady propagation
solution $\phi_0(z)$ with $z=x-c_0t$ satisfies
\begin{equation}
-c_0\partial_z \phi_0(z)=-\Gamma[f'(\phi_0) - \kappa \partial_{z}^2 \phi_0],
\label{ss-sol}
\end{equation}
where $c_0$ is the steady propagation velocity of the interface.
Mathematically, (\ref{ss-sol}) is a non-linear eigenvalue problem
for the solution $\phi_0$ with a special value of $c_0$.
Thus, $\phi_0(z)$ and $c_0$ are simultaneously
determined. The explicit form of the solution is not generally written, but we can easily confirm the following relation in the limit
$L \to \infty$ \cite{Pomeau}:
\begin{equation}
c_0 = \Gamma_{\rm int} (f(\phi_2) - f(\phi_1))
\label{formula-0}
\end{equation}
with
\begin{equation}
\Gamma_{\rm int}= \frac{\Gamma}
{ \int_{-\infty}^{\infty} dz ( \partial_{z}\phi_0(z) )^2 }.
\label{mobility}
\end{equation}  
The relation (\ref{formula-0}) 
indicates that the free energy density difference between the two local minima
drives the interface to decrease the total free energy.  The mobility of the interface is then
given by (\ref{mobility}).

When $T >0$, the noise modifies the propagation
velocity. To extract this effect clearly, we study the case $c_0=0$,
which holds when $f(\phi_2) =f(\phi_1)$. We then consider the weak noise limit $T \to 0$ ignoring nucleation events in the bulk.  
Let  $\theta(y,t)$ be
the $x$-coordinates of the interface at time $t$. The expectation of
the fluctuating quantity $\theta(y,t)$ approaches a steady
propagating state expressed as 
\begin{equation}
\bra \theta(y,t) \ket_\ss = ct+{\rm const}. 
\end{equation}
A finite value of $c$ was reported for the case $d=1$ \cite{Costantini2001},
in which the nature of the driving force was found to be entropic. That is,
when the fluctuation intensity around $\phi=\phi_1$ is larger than that
around $\phi=\phi_2$, the entropy density in the region with 
$\phi=\phi_1$ is larger and then
the region of $\phi=\phi_1$ becomes larger, leading to $c>0$.  The
formula for $c$ is expressed as \cite{Costantini2001}
\begin{eqnarray}
  c=\Gamma_{\rm int} 
 \frac{T}{2}
 \left( \frac{1}{\xi_2} - \frac{1}{\xi_1} \right) +O(T^{\frac{3}{2}}),
\label{1d-result}
\end{eqnarray}
where  $\xi_{i} \equiv \sqrt{\kappa/f''(\phi_i)}$ is the correlation
length of fluctuations in the bulk region with $\phi=\phi_i$.
It should be noted that (\ref{1d-result}) was  confirmed by
numerical simulations \cite{Costantini2001}.


In this Letter, we study $c$ for the system in two dimensions.  
Because the driving force of the interface motion is entropic, we expect 
that an entropic contribution $s(\phi)$ to the macroscopic free energy
plays an essential role in the determination of $c$. We then conjecture
\begin{equation}
c=\Gamma_{\rm int} [- T (s(\phi_2)-s(\phi_1)) ] +O(T^{\frac{3}{2}}),
\label{formula-T}
\end{equation}
which means that the difference between the entropic contributions in each
bulk region leads to the driving force. The question now is whether
or not $c$ can be expressed in such a form. Even if the form of (\ref{formula-T})
is correct,  the functional form of $s(\phi)$ is not immediately obtained from
the model (\ref{model}). We thus
need to derive $c$ for the system in two dimensions. However, because 
the derivation method in Ref. \cite{Costantini2001}, which follows
the method proposed in Refs. \cite{Currie1980, DeLeonardis1980, Cattuto2001}, 
is specific to the one-dimensional case, we have to develop a
general method of deriving $c$.

{\it Main result.---}
We derive the stochastic interface dynamics from the stochastic model
(\ref{model}). As far as deterministic systems are concerned, there have been many methods used to
derive the equation for interface motion \cite{Kuramoto1980, Kuramoto1989, Meron1992, Cross1993, Ohta1994, Karma1997, Hiraizumi2021}.
The essence of these methods is to extract
the interface motion as the slowest dynamics while separating other fast
variables. 
We generalize the methods above to analyze stochastic systems in one and higher
dimensions. We then obtain the formula (\ref{formula-T}) with
\begin{equation}
  s(\phi_i)= -\frac{1}{2}\int_{|p|\le k_{c}}\frac{dp}{2\pi}
  \frac{\xi_i^{-2}-p^2}{p^2+\xi_i^{-2}}
\label{s-1d}
\end{equation}
in one dimension and
\begin{equation}
s(\phi_i)= -\frac{1}{2}\int_{|\bv{p}|\le k_{c}}\frac{d^2\bv{p}}{(2\pi)^2}
  \frac{\xi_i^{-2}}{|\bv{p}|^2+\xi_i^{-2}}
\label{s-2d}  
\end{equation}
in two dimensions. The right-hand side of (\ref{s-1d})
is calculated as
\begin{equation}
  s(\phi_i)=  -\frac{1}{2\pi}
  \left[ \frac{2}{\xi_i}\tan^{-1}(\xi_i k_c)-k_c \right] .
\label{s-1d-ex}
\end{equation}  
By substituting this result  into (\ref{formula-T}) and
taking the limit $k_c \to \infty$, we obtain (\ref{1d-result}). 
Then, for the two-dimensional case,
the right-hand side of (\ref{s-2d}) is calculated as 
\begin{equation}
  s(\phi_i)=  -\frac{1}{8 \pi \xi_i^2}  \ln(\xi_i^2 k_c^2+1),
\label{s-2d-ex}
\end{equation}
where the cut-off wavenumber $k_c$ should remain finite.
This means that the stationary propagation velocity for
the model with $d=2$ singularly depends on
the ultra-violet cut-off $k_c$. In other words, we need to
specify a value of the cut-off $k_c$ to study a measurement
result of the propagating velocity.



\begin{figure}[t]
\begin{center}
\includegraphics[width=80mm]{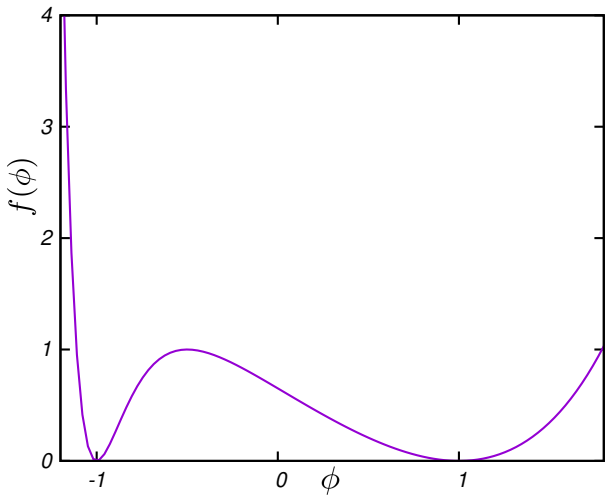}
\caption{Shape of the mesoscopic  free energy density \eqref{example-f}
with $b_1=0.5$, $b_2=5.0$, and $\phi_0=-0.5$.}
\label{fig1}
\end{center}
\end{figure}

{\it Numerical simulations.---}
Because the  cut-off dependence of the formula (\ref{s-2d-ex})
is rather striking, we now confirm this result using numerical simulations.
We note that $s(\phi_1)=s(\phi_2)$ when $f''(\phi_1)=f''(\phi_2)$.
Therefore, an asymmetric landscape of $f(\phi)$ is necessary for
the appearance of the entropic driving force. On the basis of this fact,
we assume the local free energy density $f(\phi)$ is given by
\small
\begin{eqnarray}
  f(\phi)=\biggl(\frac{1-\exp[b_1(\phi-1)]}
  {1-\exp[b_1(\phi_0-1)]}\frac{1-\exp[-b_2(\phi+1)]}{1-\exp[-b_2(\phi_0+1)]}
  \biggr)^2,
\label{example-f}
\end{eqnarray}
\normalsize
as in Ref. \cite{Costantini2001}.
Here, $f$ satisfies the condition
that $\phi_1=-1,\phi_2=1$ and $f(\phi_1)=f(\phi_2)$. In Fig. \ref{fig1}, we show the form of the local free energy density $f$. It can be seen that
the potential is highly asymmetric, i.e.,  $f''(\phi_1)\gg f''(\phi_2)$.
Examples of asymmetric free energy density $f$ were presented in Refs. \cite{Boukheddaden2005,Miyashita2005}.



\begin{figure}[t]
\includegraphics[width=80mm]{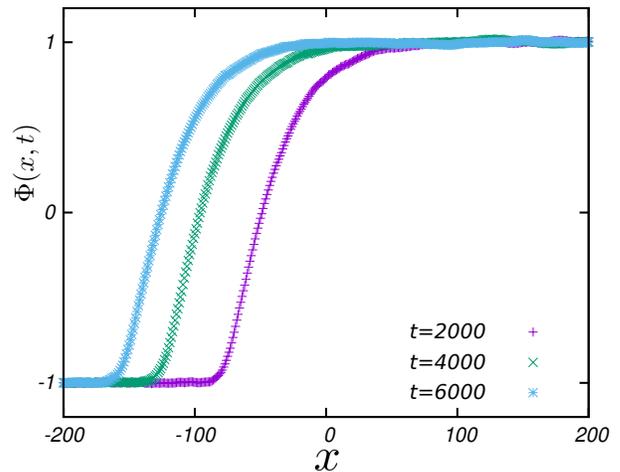}
\caption{
Time evolution of the pattern averaged in the $y$-direction.
The free energy density $f(\phi)$ is the same as that in Fig. \ref{fig1}. 
The other parameter values are $\kappa=1600$, $\Gamma=0.1$,  $T=0.5$,
$L=400$ and $L_y=100$. Noting that $\xi_1=4.33$ and $\xi_2=27.3$ for these
parameters, we choose $\Delta x=1.0$ and $\Delta t=5.0\times10^{-4}$ \cite{SM}.}
\label{fig2}
\end{figure}

\begin{figure}[t]
\begin{center}
\includegraphics[width=80mm]{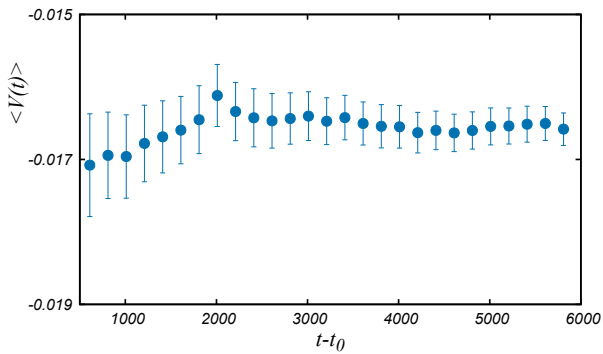}
\caption{$\langle V(t) \rangle$ as a function of $t-t_0$.
The parameter values are the same as those in Fig. \ref{fig2}.
$t_0=50$. Eighty samples are used to estimate $\langle V(t) \rangle$ with
error-bars.}
\label{fig3}
\end{center}
\end{figure}

We define a discrete model on a square lattice with a spatial mesh size of $\Delta x$, considering that $\Delta x$ should be smaller than $\xi_1$ and $\xi_2$. 
The discrete model is obtained by discretizing (\ref{model}), where $\partial_x^2+\partial_y^2$ is replaced by the finite difference Laplacian.
For the initial condition
\begin{equation}
\phi(\bv{r},0)=
\frac{\phi_2-\phi_1}{2} \tanh\left(\frac{x}{40}\right)
+ \frac{\phi_1+\phi_2}{2},
\end{equation}
the stochastic time evolution is performed using the Heun method. 
We then measure
\begin{eqnarray}
\Phi(x,t) \equiv \frac{1}{L_y}\int_0^{L_y} d y \phi(x,y,t),
\end{eqnarray}
which describes the $x$-profile averaged in the $y$-direction.
In Fig. \ref{fig2},  we show an example of
$\Phi(x, t)$ for several values of $t$. 
For the interface position $X(t)$ defined by $\Phi(X(t),t)=0$,
we define a time-averaged velocity as 
\begin{equation}
V(t)\equiv \frac{X(t)-X(t_0)}{t-t_0},
\end{equation}
where $t_0$ is chosen to be much larger than the relaxation time
to the steady propagating state.  In Fig. \ref{fig3}, we plot
the expectation value of $V(t)$, which gives the numerically estimated
value of $c$. 
We then confirm that $c$ is proportional to $T$ for $T \leq 1$ \cite{SM}.


\begin{figure}[t]
\begin{center}
\includegraphics[width=85mm]{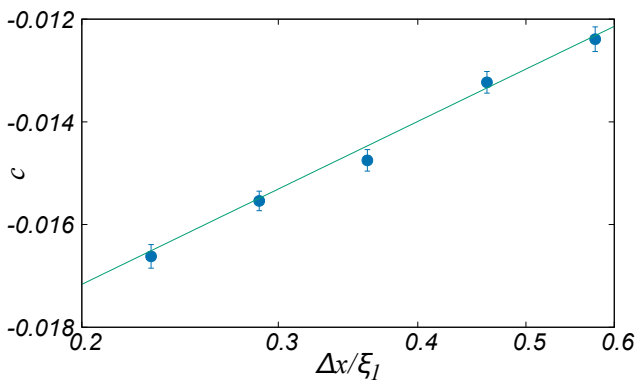}
\caption{ $\Delta x$-dependence of the propagating velocity $c$ for
the system in two dimensions. 
The parameter values are the same as those in Fig. \ref{fig2}.
The blue circles represent numerical results given by $ \bra V(t=6000) \ket $.
The green curve represents (\ref{formula-T}) with (\ref{s-2d-ex}).
We checked the validity of the values of the numerical parameters
$\Delta x$ and $\Delta t$ \cite{SM}.}
\label{fig4}
\end{center}
\end{figure}

We now perform the same calculation for systems with different
values of $\Delta x$. The results are displayed in Fig. \ref{fig4}.
It is observed that 
$c$ does not go to a definite value as $\Delta x$ becomes
smaller under the condition that $\Delta x < \xi_1 \ll \xi_2$,
which is in contrast with the one-dimensional case.
To compare the numerical data with the theoretical result,
we overlay the graph of (\ref{formula-T}) with (\ref{s-2d-ex})
in Fig. \ref{fig4}, where we choose 
\begin{equation}
  k_c=\sqrt{ \left(\frac{2\pi}{\Delta x}\right)^2+
    \left(\frac{2\pi}{\Delta x}\right)^2}
   =\frac{2\sqrt{2}\pi}{\Delta x},
\end{equation}    
which corresponds to the largest magnitude of the wave vector
in the numerical simulations. 
We find that the theoretical calculation is consistent
with the numerical simulations. We thus conjecture that
$|c|$ diverges in the limit $\Delta x \to 0$ even in
numerical simulations.


{\it Sketch of the derivation.---}
We first take the limit $L \to \infty$.
Let $T$ be sufficiently small such that 
the noise effect can be studied as a perturbation of
the deterministic system. To express this smallness,
we replace $T$ with $\epsilon^2 T$, where $\epsilon$ is
a small dimensionless parameter. 
When a perturbation is imposed on the stationary solution
$\phi_0(x)$, the response is divided into the interface
motion $\theta(y,t)$ and the rest. That is, we express
the perturbation solution as 
\begin{equation}
\phi(x,y,t) = \phi_0(z) + \epsilon\rho_1(z,y,t)+O(\epsilon^2) 
\label{p-sol}
\end{equation}
with the co-moving coordinate $z=x-\theta(y,t)$.
Assuming that $\partial_y \theta$ is small and proportional
to $\epsilon$, we introduce a large scaled coordinate $Y$ as
$Y=\epsilon y$, and define $\Theta(Y,t)=\theta(y,t)$. 
The time evolution of $\Theta(Y,t)$ is described by 
\begin{equation}
  \partial_t \Theta =\epsilon \Omega_1([\Theta])
  +\epsilon^2 \Omega_2([\Theta])+
 O(\ep^3),
\label{theta_form}
\end{equation}
where $[\Theta]$ represents the dependence of $\partial_Y \Theta$
and $\partial_Y^2 \Theta$.
By substituting (\ref{p-sol}) and (\ref{theta_form}) into (\ref{model}),
we can determine the statistical properties of  $\rho_1$, $\Omega_1$, and
$\Omega_2$. This calculation method is regarded as a generalization
of the method in Ref. \cite{Kuramoto1980} to stochastic systems. 

The stationary propagation velocity $c$ is obtained as
$c= \epsilon \bra \Omega_1 \ket_\ss+ \epsilon^2 \bra \Omega_2 \ket_\ss+O(\epsilon^3)$.
We calculate $ \bra \Omega_1 \ket_\ss=0$ and
\begin{equation}
\bra \Omega_2 \ket_\ss= \frac{\Gamma}{2}
\frac{ \int_{-\infty}^{\infty} dz
  (\partial_z \phi_0) f^{(3)}(\phi_0) \bra \rho_{1}^{2} \ket_\ss }
  { \int_{-\infty}^{\infty}dz(\partial_z \phi_0)^2 }.
\label{omega2}
\end{equation}
This formula was derived in Ref. \cite{Iwata2010}.
Now, from the time-reversal symmetry of the steady state in the co-moving frame,
we find that the integral of the numerator in (\ref{omega2}) is
expressed as $\lim_{\Lambda \to \infty}[\Psi(z=\Lambda)-\Psi(z=-\Lambda)]$
using a function $\Psi(z)$ given by 
\begin{align}
  \Psi(z)=& f^{(2)}(\phi_0(z))\bra \rho_1(z,y)^2\ket_\ss
           \nonumber \\
   &  -\kappa \left[\bra (\partial_z \rho_1(z,y))^2 \ket_\ss-
   \bra (\partial_y \rho_1(z,y))^2 \ket_\ss \right].
\end{align}
Therefore, (\ref{omega2}) takes the form (\ref{formula-T}) with
\begin{equation}
T s(\phi_i)=-\frac{1}{2} \lim_{\Lambda \to \infty} \Psi(\mu_i \Lambda) , 
\end{equation}
where $\mu_1=-1$ and $\mu_2=1$. Finally, by evaluating $\Psi(z)$,
we obtain (\ref{s-2d}). The calculation result is immediately generalized
to $d$-dimensional systems  \cite{SM}. In particular, for the case $d=1$, we obtain
(\ref{s-1d}).

{\it Concluding remarks.---}
We have derived the formula (\ref{formula-T}) with (\ref{s-2d}) 
for the entropic driving force in the mesoscopic model (\ref{model}).
Although we have studied the case $c_0=0$, it is straightforward to
derive the propagation velocity for $c_0\neq0$ as $c_0+c$, which is denoted by $c_{*}$.  
We have found that the entropic force singularly depends on the microscopic
cut-off of the mesoscopic model (\ref{model}) in two dimensions.
This discovery suggests the need for further study. 

The most important challenge is an experimental observation
of the entropic force in two or three dimensions. 
As an experimental system, we propose to study a spin-crossover complex
where high-spin and low-spin states can coexist with an
interface \cite{Boukheddaden2016}.
Since curvatures of the free energy at these states
are different, we expect the entropic contribution of the force
generating the interface propagation. Here, through the measurement
of the space-time correlation of order parameter fluctuations,
$\Gamma$ and the correlation lengths are evaluated.
By measuring $c_{*}$ directly at sufficiently low temperatures $T_1$ and $T_2$, we may estimate the value of $k_c$ by using $c_{*}(T_2)-c_{*}(T_1)\simeq -\Gamma_{\rm int} (T_2-T_1)(s(\phi_2)-s(\phi_1))$
with the assumption that in the low temperature region $f(\phi)$ does not depend on $T$.

From the theoretical viewpoint, it is significant to generalize
our formula (\ref{formula-T}) for studying various systems such as
conserving systems \cite{Chan_Hilliard} and
out-of-equilibrium systems \cite{Panja2004, Rocco2002, Lemarchand2011}. More
fundamentally, in pursuit of a microscopic understanding of the cut-off,
one can attempt to consider the derivation of the mesoscopic
description from more microscopic systems such as lattice models
\cite{Spohn1983} or Hamiltonian systems \cite{KNS}. Although there have
been many related studies since Ref. \cite{Zwanzig}, the explicit
determination of coefficients of the mesoscopic model is not easy as argued in Ref. \cite{Saito}. To develop a theory explaining
the cut-off dependence based on a microscopic description should be
another goal of non-equilibrium statistical mechanics.
As another direction of study, the universality class of
stochastic interface motion will be explored by studying the fluctuation
properties of the interface motion. With regard
to this problem, we point out that the microscopic cut-off dependence
observed in the Kardar-Parisi-Zhang equation \cite{KPZ, Takeuchi2018} comes from a non-linear term
that is not relevant in our problem \cite{SM}. Thus,  the microscopic
cut-off dependence reported in this Letter has never been studied so far.


The authors thank K. Kanazawa and K. Saito for useful comments. 
This study was supported by JSPS KAKENHI (Grant Numbers JP19H05795,
JP20K20425, and JP22H01144).




\clearpage
\onecolumngrid

\setcounter{equation}{0}
\renewcommand{\theequation}{S\arabic{equation}}

\setcounter{figure}{0}
\renewcommand{\thefigure}{S\arabic{figure}}

\begin{center}
{\large \bf Supplemental Material:  \protect \\ 
Microscopic cut-off dependence of an entropic force in interface propagation of stochastic order parameter dynamics }\\
\vspace*{0.3cm}
Yutaro Kado and Shin-ichi Sasa
\end{center}


In the first part of Supplemental Material (SM),
we derive (11) following the sketch of
derivation in the main text.  In the second part of
SM, we present details of numerical simulations.

\section{I. Derivation of the main result}

In Sec. I A, we briefly explain
a setup of perturbation theory. In Sec. I B,
we introduce some quantities that are useful in the analysis. 
In Sec. I C, we review the derivation of (21).
The arguments to this point are well-known. Section I D
is the technical highlight, which has not been reported in
literature, as far as we know.  Using the time-reversal symmetry,
we derive (22). In Sec. I E, we obtain (11) by calculating
fluctuations.  In Sec. I F,  we address some remarks on
the theoretical calculation.
Throughout the first part of SM,
we set $\Gamma=1$ without loss of generality.

\subsection{A. Setup}
\label{setup}

We rewrite (2) in the main text as 
\begin{eqnarray}
 \partial_{t}\phi=-f'(\phi_0(x))
 + \kappa (\partial_{x}^2 + \partial_{y}^2) \phi +  \eta.
\label{dynamics_2dim}
\end{eqnarray}
Since we take the limit $L \to \infty$, the boundary
conditions in the $x$-direction become
$\phi(\bv{r}) \to \phi_1$ as $x \to -\infty$ and
$\phi(\bv{r}) \to \phi_2$ as $x \to \infty$. 
We study the case $c_0=0$ in (4) in the main text. That is,
we assume that there exists $\phi_0(x)$ satisfying 
\begin{eqnarray}
0=-f'(\phi_0) + \kappa \partial_x^2 \phi_0.
\label{T=0}
\end{eqnarray}
To express the smallness of noise intensity,
we replace $T$ by  $\epsilon^2 T$, where  $\epsilon$ is a
small dimensionless parameter. Thus, $\eta$ satisfies
\begin{equation}
  \bra \eta(x,y,t) \eta(x',y',t') \ket=2T\epsilon^2
  \delta(x-x')\delta(y-y')\delta(t-t').
\label{noise}
\end{equation}  
As presented in (19) in the main text,
we express a perturbation solution for $\phi_0$ as 
\begin{eqnarray}
\phi(x,y,t) =\phi_0(z)  + \epsilon \rho_{1}(z,y,t)
+ \epsilon^{2} \rho_{2}(z,y,t) + O(\epsilon^3)
\label{expansion_form_2dim}
\end{eqnarray}
with the co-moving coordinate  $z\equiv x-\theta(y,t)\in[-\infty, \infty]$,
where $\theta(y,t)$ represents an interface position for each $y$ at time $t$. 
Assuming that $\partial_y \theta$ is small and proportional to $\epsilon$,
we introduce a large scaled coordinate $Y=\epsilon y$ and define
$\Theta(Y,t) \equiv \theta(y,t)$. The time evolution of $\Theta(Y,t)$ is
described by 
\begin{eqnarray}
\partial_t \Theta(Y,t)=\epsilon \Omega_{1}([\Theta]) + \epsilon^{2} \Omega_{2}([\Theta])+ O(\epsilon^3),
\label{theta_form_2dim}
\end{eqnarray}
where $[\Theta]$ represent the dependence of $\partial_Y \Theta$ and 
$\partial_Y^2 \Theta$. We substitute (\ref{expansion_form_2dim})
and (\ref{theta_form_2dim}) into (\ref{dynamics_2dim}). In this calculation,
we note 
\begin{equation}
  \partial_t \phi=-\partial_t \Theta(\partial_z\phi_0+\partial_z \rho)
  +\partial_t \rho,
\end{equation}  
with $\rho=\epsilon \rho_1+\epsilon^2\rho_2+O(\epsilon^3)$.
We then collect
terms proportional to $\epsilon$. From this, we can determine
$\rho_1$ and $\Omega_1$. Next, collecting terms proportional to
$\epsilon^2$, we derive $\Omega_2$. The propagation velocity $c$
is given by
\begin{equation}
c= \epsilon \bra \Omega_1 \ket_{\ss} +\epsilon^2 \bra \Omega_2 \ket_{\ss} + O(\epsilon^3).
\label{c-Omega}
\end{equation}
Below, we explain these calculations.

\subsection{B. Preliminaries}
\label{preliminary}

For later convenience,
we introduce some useful quantities, a linear operator $\hat L_z$,
projection operators $\hat P$ and $\hat Q$, noises $\tilde \eta(z,y,t)$,
$\hat P \tilde \eta(z,t)$, and $\hat Q \tilde \eta(z,y,t)$.

\subsubsection{1. Operators}
We first define the linear operator 
\begin{eqnarray}
\hat{L}_z\equiv- f''(\phi_0(z)) +\kappa \partial_{z}^2 ,
\end{eqnarray}
which describes the linear dynamics of perturbations
around $\phi_0(z)$.  Setting $u_0(z)\equiv \partial_z \phi_0(z)$,
we can confirm
\begin{equation}
\hat L_z u_0=0
\end{equation}
from the space-translation symmetry of the system.
Formally, $u_0$ corresponds to the Goldstone mode associated
with the translation symmetry breaking by the solution $\phi_0$.
Physically,
$u_0$ describes the interface mode.

We next define the projection to the  mode $u_0$ by
\begin{align}
\hat{P} g  \equiv  \frac{(u_0,g)}{(u_0,u_0)} u_0
\end{align}
for $g(z,y)$, where $(g_1,g_2)$ represents the inner product given by
\begin{equation}
(g_1,g_2)\equiv\frac{1}{L_y}
\int_0^{L_y} dy \int_{-\infty}^{\infty} dz g_1(z, y)g_2(z,y)
\end{equation}
for functions $g_1(z,y)$ and $g_2(z,y)$ defined on
$D\equiv [-\infty,\infty] \times [0, L_y]$.
We also define $\hat Q\equiv 1-\hat P$.  

\subsubsection{2. Noises}
For a given noise $\eta(x,y,t)$, we define
\begin{equation}
\tilde \eta(z,y,t)\equiv \epsilon^{-1}\eta(z+\theta(y,t),y,t).  
\end{equation}
Statistical properties of $\tilde\eta$ are characterized
by $\bra \tilde \eta \ket$=0 and
\begin{equation}
\bra \tilde\eta(z,y,t) \tilde\eta(z',y',t') \ket=
2T \delta(z-z') \delta(y-y') \delta(t-t')  .
\label{tilde-eta-int}
\end{equation}
For this noise $\tilde\eta$, we consider the projected noise
$\hat P \tilde\eta(z,t)$. The noise intensity
of $\hat P \tilde\eta$  is calculated as
\begin{align}
 \bra \hat P \tilde \eta(z,t)\hat P \tilde\eta(z',t') \ket  
=\bra \tilde \eta(z,y,t)\hat P \tilde\eta(z',t') \ket  
= 2T\frac{u_0(z)u_0(z')}{(u_0,u_0) L_y}  \delta(t-t').
\label{P}
\end{align}
From this result, we immediately obtain
\begin{equation}
  \bra \hat Q \tilde\eta(z,y,t)\hat Q \tilde\eta(z',y',t')
  \ket  
=\bra \tilde\eta(z,y,t)\hat Q \tilde\eta(z',y',t') \ket 
=
2T\left[
  \delta(z-z')\delta(y-y')-\frac{u_0(z)u_0(z')}{(u_0,u_0) L_y}
\right] \delta(t-t').
\label{Q}
\end{equation}

\subsection{C. Derivation of (21)}
\label{der-21}

In this section, we derive (21) in the main text.
As described in Sec. I C, we substitute 
(\ref{expansion_form_2dim}) and (\ref{theta_form_2dim})
into (\ref{dynamics_2dim}). Then, collecting terms
proportional to $\epsilon$, we obtain
\begin{equation}
(\partial_t - \hat{L}_z - \kappa \partial_{y}^2)
  \rho_{1} =B_1
\label{rho_1_2dim}
\end{equation}
with
\begin{equation}
B_1(z,y,t)=    \Omega_{1}([\Theta]) u_0(z) + \tilde\eta(z,y,t).
\end{equation}
Because $(\partial_t - \hat{L}_z - \kappa \partial_{y}^2) u_0=0$,
there exists a solution $\rho_1$ to the linear equation (\ref{rho_1_2dim})
only when $(u_0, B_1)=0$. To have a consistent perturbation solution,
we impose this condition, which is referred to as the solvability
condition.  The solvability condition leads to
\begin{equation}
\Omega_{1}([\Theta])= - \frac{(u_0,\tilde\eta)}{(u_0,u_0)}.
\label{omega_1_2dim}
\end{equation}
This gives $\bra \Omega_1 \ket_\ss =0$.
Then,  (\ref{rho_1_2dim}) becomes
\begin{equation}
 (\partial_t - \hat{L}_z - \kappa \partial_{y}^2) \rho_{1}
= \hat{Q} \tilde\eta.
\label{rho_1_1_2dim}
\end{equation}
To have the unique solution of $\rho_1$, we further impose 
$(u_0,\rho_1)=0$.

Next, collecting terms proportional to $\epsilon^2$, we obtain
\begin{equation}
(\partial_t - \hat{L}_z - \kappa \partial_{y}^2) \rho_{2} =B_2
\label{rho_2_2dim}
\end{equation}
with
\begin{equation}
B_2=  \Omega_{2}([\Theta]) u_0 +\Omega_{1}([\Theta]) \partial_z \rho_{1} 
- \frac{1}{2}f^{(3)}(\phi_0) \rho_{1}^{2} - \kappa (\partial_{Y}^2 \Theta) u_0(z) + \kappa (\partial_{Y} \Theta)^2 \partial_z u_0
-2 \kappa (\partial_Y \Theta) \partial_z \partial_y \rho_1.
\end{equation}
We then impose the solvability condition $(u_0,B_2)=0$,
which is expressed as
\begin{equation}
  \Omega_{2}([\Theta]) = \kappa \partial_Y^2 \Theta
  + \frac{1}{(u_0,u_0)}
  \left[
    \left(u_0, \frac{1}{2} f^{(3)}(\phi_0) \rho_{1}^{2} \right) -
    \Omega_1 (u_0, \partial_z \rho_{1})
  \right],
\label{omega_2_2dim}
\end{equation}
where we have used $(u_0, \partial_{ z } u_0)=0$ and
$(u_0,  \partial_z \partial_y \rho_1)=0$.

Now, we calculate $\bra \Omega_2 \ket_\ss$. We first notice the relation
\begin{equation}
\bra (\hat P \tilde\eta, \partial_z \rho_1) \ket=0  .
\label{noise-product}
\end{equation}
This is obtained from the direct calculation as 
\begin{align}
 & (u_0,u_0)L_y
  \int_{-\infty}^\infty  dz
  \int_0^{L_y}  dy
  \bra \hat P \tilde\eta(z,t) \partial_z \rho_1(z,y,t) \ket
  \nonumber \\
=&
\int_{-\infty}^\infty dz\int_0^{L_y} dy \int_{-\infty}^\infty  dz'
\int_0^{L_y} dy' u_0(z')u_0(z)
\bra \tilde\eta(z',y',t) \partial_z \rho_1(z,y,t) \ket   \nonumber \\
=& 
T\int_{-\infty}^\infty dz\int_0^{L_y} dy \int_{-\infty}^\infty  dz'
\int_0^{L_y} dy' u_0(z')u_0(z)
\partial_z
\left[\delta(z-z')\delta(y-y')-\frac{u_0(z)u_0(z')}{(u_0,u_0)L_y} \right]
\nonumber \\
=&0.
\end{align}
Here, to get the third line, we have used (\ref{Q}) with noting
the Stratonovich rule of the multiplication, and to get the last
line we have used $ \int_{-\infty}^{\infty} dz u_0(z) \partial_z u_0(z) =0$.

Next, we take the expectation of (\ref{omega_2_2dim}) in the steady state.
Noting the relation
(\ref{noise-product}), we have 
\begin{equation}
\langle \Omega_1([\Theta]) (u_0, \partial_z \rho_{1}) \rangle = 0.
\label{omega_2_1_2dim}
\end{equation}
Using this result, we obtain 
\begin{equation}
\bra  \Omega_2 \ket_\ss 
= \frac{ \int_{-\infty}^{\infty } dz u_0(z) f^{(3)}(\phi_0(z) )
 \langle \rho_{1}(z,y)^{2} \rangle_\ss }{2\int_{-\infty}^{\infty } dz u_0(z)^2 } , 
\label{eq21}
\end{equation}
where we have used the fact that $\langle \rho_{1}(z,y)^{2} \rangle_\ss$
does not depend on $y$. This is (21) in the main text.

\subsection{D. Derivation of (22)}
\label{der-22}

In this section, we derive (22) in the main text. First, noting
the trivial relation 
\begin{equation}
  u_0(z) f^{(3)}(\phi_0(z))
  =\frac{d\phi_0}{dz} \frac{df^{(2)}(\phi_0)}{d\phi_0}
  =\frac{df^{(2)}(\phi_0(z))}{dz},
\end{equation}
we obtain
\begin{equation}
  \int_{-\infty}^{\infty} dz \frac{1}{2}
  u_0f^{(3)}(\phi_0)\rho_{1}^{2}
  =\frac{1}{2}
  \left.
  \left[
    f^{(2)}(\phi_0) \rho_{1}^{2}   - \kappa ( \partial_z \rho_{1} )^2
    \right]
  \right\vert_{z=-\infty}^{z=\infty}
 + \int_{-\infty}^{\infty} dz \partial_z \rho_1\hat{L}_z \rho_1 .
\label{partial_2dim}
\end{equation}
Here, from the time-reversal symmetry of the steady state in the
co-moving frame, a non-trivial relation
\begin{eqnarray}
  \int_{-\infty}^{\infty} dz \langle \partial_z
  \rho_{1}(z,y)\hat{L}_z \rho_{1}(z,y)  \rangle_\ss
  = \left.
  \frac{\kappa}{2} \langle
  (\partial_y \rho_{1}(z,y) )^2 \rangle_\ss
  \right|_{z=-\infty}^{z=\infty}
\label{time_reversal_2dim}
\end{eqnarray}
holds. The proof will be given in the next paragraph. Then, 
substituting (\ref{time_reversal_2dim}) into
(\ref{partial_2dim}),  we obtain 
\begin{equation}
\bra \Omega_2 \ket
=
\frac{ \Psi(z=\infty)-\Psi(z=-\infty) }
{2\int_{-\infty}^{\infty } dz u_0(z)^2 } 
\label{c-Psi}
\end{equation}
with
\begin{equation}
\Psi(z) \equiv f^{(2)}(\phi_0(z)) \langle \rho_{1}(z,y)^{2} \rangle_\ss  - \kappa[ \langle( \partial_z \rho_{1}(z,y) )^2 \rangle_\ss- \langle( \partial_y \rho_{1} (z,y))^2 \rangle_\ss ] .
\label{psi-def}
\end{equation}
This is (22) in the main text. Recalling (\ref{c-Omega}),
we compare (\ref{c-Psi}) with (9) in the main text. Setting 
\begin{align}
  Ts(\phi_2) &= -  \frac{1}{2}
  \lim_{\Lambda \to \infty} \Psi(z=\Lambda),  \label{s2}\\
  Ts(\phi_1) &= - \frac{1}{2}
  \lim_{\Lambda \to \infty} \Psi(z=-\Lambda), \label{s1}
\end{align}
we confirm the validity of the expression of (9).

We show a proof of the key relation (\ref{time_reversal_2dim}).
First, from the time-reversal symmetry of fluctuations of $\rho(z,y,t)$,
we have
\begin{equation}
  \bra \partial_z\rho_1(z,y,t) \rho_1(z,y,t+t') \ket_\ss
=  \bra \partial_z\rho_1(z,y,t) \rho_1(z,y,t-t') \ket_\ss
\end{equation}
for any $t$ and $t'$. Taking the limit $t'\to 0$, we find 
\begin{equation}
\bra \partial_z\rho_1(z,y,t) \partial_t \rho_1(z,y,t) \ket_\ss =0. 
\label{4_b_2}
\end{equation}
By repeating the calculation for (\ref{noise-product}), we can derive
\begin{eqnarray}
  \int_{-\infty} ^{\infty} dz
  \langle \partial_z \rho_1(z,y,t) \hat{Q}
  \tilde\eta(z,y,t) \rangle_\ss = 0.
\label{4_b_4}
\end{eqnarray} 
Now, we multiply (\ref{rho_1_1_2dim}) by $\partial_z \rho(z,t)$
and integrate over $[-\infty,\infty]$ with respect to $z$. Using
(\ref{4_b_2}) and (\ref{4_b_4}),  we obtain
\begin{align}
  \int_{-\infty} ^{\infty} dz \langle \partial_z \rho_{1}(z,y,t) \hat{L}_z \rho_{1}(z,y,t) \rangle_\ss
&= - \kappa \int_{-\infty} ^{\infty} dz  \langle
  \partial_z \rho_{1}(z,y,t) \partial_y^2 \rho_{1}(z,y,t)
  \rangle_\ss  \nonumber \\
&=  \kappa \int_{-\infty} ^{\infty} dz \langle
  \partial_z \partial_y\rho_{1}(z,y,t) \partial_y \rho_{1}(z,y,t)
  \rangle_\ss  \nonumber \\
  &= \left. \frac{\kappa}{2}
    \langle
   (\partial_y\rho_{1}(z,y,t))^2 \rangle_\ss \right|_{z=-\infty}^{z=\infty},
\end{align} 
which is  (\ref{time_reversal_2dim}).

\subsection{E. Derivation of (11)}
\label{der-11}

We set $\mu_1=-1$ and $\mu_2=1$. In this section,
we evaluate $\lim_{\Lambda \to \infty}\Psi(z=\mu_i \Lambda)$
which is connected to $s(\phi_i)$ by (\ref{s1}) and (\ref{s2}).
We fix a sufficiently large $\Lambda$. It is reasonable to
assume that fluctuations at a point $(z,y)=(\mu_i \Lambda, y_0)$
are equivalent to those at a point of the homogeneous system
with $\bra \phi \ket=\phi_i$ without any interfaces, because the
point $(\mu_i \Lambda, y_0)$ is far from the interface.
With this assumption, it is easy to analyze (\ref{rho_1_1_2dim}).
We have the following estimations:
\begin{align}
  \langle \rho_1(\mu_i \Lambda, y_0)^2
  \rangle_\ss
&= \int_{|\bm{p}| \leq k_c }
\frac{d^2 \bm{p}}{(2 \pi)^2} \frac{T}{ f^{(2)}(\phi_i) + \kappa |\bm{p}|^2 }, \\
\langle (\partial_z\rho_1(\mu_i \Lambda, y_0)^2 \rangle_\ss  
&=
\int_{|\bm{p}|\leq k_c} \frac{d^2 \bm{p} }{(2 \pi)^2}
\frac{T p_x^2}{ f^{(2)}(\phi_i) + \kappa |\bm{p}|^2 },  \\
\langle (\partial_y\rho_1(\mu_i \Lambda, y_0) )^2 \rangle_\ss
&=
\int_{ |\bm{p}|\leq k_c} \frac{d^2 \bm{p} }{(2 \pi)^2}
\frac{T p_y^2}{ f^{(2)}(\phi_i) + \kappa |\bm{p}|^2 },
\end{align}
where $\bm{p}=(p_x,p_y)$.
It should be noted that the results are independent of $\Lambda$.
Here, by substituting $f^{(2)}(\phi_i)=\kappa/\xi_i^{2}$ into these
equations and  combining them as the form (\ref{psi-def}), 
we obtain 
\begin{align}
\Psi(z=\mu_i \Lambda) &= T 
\int_{|\bm{p}| \leq k_c} \frac{ d^2 \bm{p} }{ (2\pi)^2 }
  \frac{\xi_{i}^{-2}-p_x^2+p_y^2}
       { |\bm{p}|^2 +\xi_{i}^{-2} }, \nonumber  \\
&=\frac{T  }{4\pi \xi_i^{2}} {\rm ln} (\xi_i^2k_c^2+1).
\end{align}
By substituting them into (\ref{s1}) and (\ref{s2}),
we obtain (11) and (12) in the main text. 

\subsection{F. Remarks}
\label{remark}

\subsubsection{1. Mobility}

To numerically calculate the right-hand side of (9) in the main text,
we have to estimate the value of the right-hand side of (6). 
Using (\ref{T=0}), we first obtain
\begin{equation}
\frac{\kappa}{2} (\partial_z \phi_0)^2 -f(\phi_0)=0,  
\end{equation}
where we have used 
$f(\phi_1)=0$, $f(\phi_2)=0$, and $\partial_z \phi_0(\pm\infty)=0$.
This result leads to 
\begin{equation}
  \int dz (\partial_z \phi_0)^2 =
\sqrt{\frac{2}{\kappa}} \int_{\phi_1}^{\phi_2} d\phi \sqrt{f(\phi)}.
\label{mobility}
\end{equation}
The right-hand side of (\ref{mobility}) is numerically estimated as
$1.80/\sqrt{\kappa}$ for the local free energy density displayed
in Fig. 1 of the main text.

\subsubsection{2. Generalization}
It is straightforward to generalize
the above arguments to the $d$-dimensional system.  We can calculate 
\begin{equation}
  s(\phi_i)=
  -\frac{1}{2} \int_{ |\bm{p}| \leq k_c} \frac{ d^d \bm{p} }{ (2\pi)^d}
  \frac{\xi_{i}^{-2} -p_1^2+ \sum_{\ell=2}^d p_\ell^2}
       { |\bm{p}|^2 +\xi_{i}^{-2} },
\end{equation}
where $\bm{p}=(p_1,\cdots,p_d)$.
In particular, for the one-dimensional system, we have
\begin{align}
  s(\phi_i)&=
  -\frac{1}{2} \int_{|p| \leq k_c} \frac{ d p }{ 2\pi}
  \frac{\xi_{i}^{-2} -p^2}
       { p^2 +\xi_{i}^{-2} },   \\
       &=-\frac{1}{2\pi}
       \left[\frac{2}{\xi_j} {\rm tan}^{-1} (\xi_j k_c) - k_c \right].
\end{align}
These are  (10) and (12) in the main text. 
For the three-dimensional system, we have
\begin{equation}
  s(\phi_i)=-\frac{1}{6\pi ^2} \left[ \frac{k_c}{\xi_i^2} - \frac{1}{\xi_i^3}  {\rm tan}^{-1} (\xi_j k_c) \right] - \frac{1}{36 \pi^2}k_c^3.
\end{equation}

\subsubsection{3. Previous Studies}

The calculation method presented in this paper
is regarded as a stochastic generalization of the method
used in the analysis of deterministic equations \cite{Kuramoto1980_ver2}.
The formulation was reported in Ref. \cite{Iwata2010_ver2}. 
For the one-dimensional system, the result of $c$ is the same as
that obtained in the previous study \cite{Costantini2001_ver2}, although
their calculation method is different from ours. 
The advantage of our method is that we can easily analyze the
$d$-dimensional system. 
We also note that our result (\ref{c-Psi}) with (\ref{psi-def})
can be obtained by using an expression for $c$ reported in
Ref. \cite{Kawasaki1982_ver2}. However, this expression is not correctly
derived. Specifically, we should prove the non-trivial relation
(\ref{time_reversal_2dim}) to derive the expression, but there
are no arguments for the proof in Ref. \cite{Kawasaki1982_ver2}. 

\subsubsection{4. Universality Class}

A one-dimensional growing interface $x=\theta(y,t)$ in a two-dimension space
$(x,y)$ is often described by the Kardar-Parisi-Zhang (KPZ) equation \cite{KPZ_ver2, Takeuchi2018_ver2}
\begin{equation}
  \partial_t \theta(y,t) = v + \nu \partial_y^2\theta
  + \lambda (\partial_y \theta)^2+ \sqrt{2D} \eta(y,t) 
\end{equation}
with the Gaussian white noise $\eta$ satisfying 
\begin{equation}
\langle \eta(y,t) \eta(y',t') \rangle = \delta(y-y') \delta(t-t').
\end{equation}
The average drift velocity is then given by
\begin{equation}
  c= v+  \lambda \bra (\partial_y \theta)^2 \ket.
\end{equation}
From the fluctuation property of the KPZ equation, it can be
easily seen that $\bra (\partial_y \theta)^2 \ket$ shows a ultra-violet
divergence. 
However, in the present problem, $\lambda=0$ in the lower order of
the perturbation theory, as discussed in Sec. I C.
Indeed, according to our perturbation theory, we have $v=O(T)$, 
$\lambda=O(T)$, $\nu=\kappa=O(1)$, and $D=O(T)$. We thus confirm
that $\lambda \bra (\partial_y \theta)^2 \ket=O(T^2)$. 
Thus, as far as we focus on the region $c=O(T)$, the non-linear term
$\lambda \bra (\partial_y \theta)^2 \ket$ is not relevant. 
Our discovery is that $v=O(T)$ exhibits the ultra-violet divergence
which is qualitatively different from the divergence observed in the KPZ
equation. It is an important problem to
clarify the universality class of our model.

\section{II. Details of Numerical Simulations}
\label{numerical}

In Fig. 4 of the main text, we report the numerical result of the propagating velocity $c$ in the two-dimensional system. In this section, we explain
the details of the numerical simulations. In Sec. II A,
we first introduce the numerical scheme we employed.
In Sec. II B, we show the numerical result for
the one-dimensional system to demonstrate that the parameter
values chosen in the main text are appropriate. In Sec. II C,
we show supporting data for the results presented in the main text.

\subsection{A. Numerical scheme}
\label{Numerical}

We discretize $(x,y)$ as $(i\Delta x,j\Delta x)$, where $i$ and $j$ are
integers satisfying $-N_x \leq i\leq N_x$ and $0\leq j<N_y$.
We note that $N_x \Delta x=L$ and $N_y \Delta x=L_y$,
where $\Delta x$ is a space interval. Let $\hat \phi(i,j)$ be defined
by $\phi(i\Delta x,j\Delta x)$. We define the discrete Laplacian as 
\begin{eqnarray}
  \Delta_2 \hat \phi(i,j)
  \equiv \frac{1}{(\Delta x)^2}
         [\hat\phi(i+1,j)+\hat\phi(i,j+1) - 4\hat\phi(i,j)
           + \hat \phi(i-1,j) + \hat\phi(i, j-1)],
\end{eqnarray}
where we impose the periodic boundary conditions in the $y$-direction
and the Dirichlet boundary conditions in the $x$-direction.

We then  discretize time $t$ as $n\Delta t$, where $n$ is a non-negative
integer and $\Delta t$ is a time interval. 
Let $\hat{\phi}(i,j,n)$ represent $\phi(i\Delta x,j\Delta x, n\Delta t)$.
The time evolution of
$\hat \phi$ is given by the Heun method defined by
\begin{eqnarray}
\hat \phi(i,j,n+1)\equiv \hat \phi(i,j,n) + \frac{1}{2}(h_1(i,j,n)+h_2(i,j,n)),
\end{eqnarray}
where $h_1(i,j,n) $ and $h_2(i,j,n)$ are given by 
\begin{equation}
  h_1(i,j,n) \equiv \Gamma[-f'(\hat \phi(i,j,n))+
    \kappa \Delta_2 \hat \phi(i,j,n)] \Delta t
  + \sqrt{2\Gamma T} \frac {\sqrt{\Delta t}}{\Delta x } \eta(i,j,n), 
\end{equation}
and 
\begin{equation}
h_2(i,j,n) \equiv \Gamma[-f'(\hat \phi'(i,j,n))+ \kappa \Delta_2 \hat \phi'(i,j,n)] \Delta t+ \sqrt{2\Gamma T} \frac {\sqrt{\Delta t}}{\Delta x } \eta(i,j,n)
\end{equation}
with
\begin{equation}
\hat \phi'(i,j,n) \equiv \hat \phi(i,j,n) + h_1(i,j,n).
\end{equation}
Here, $\eta(i,j,n)$ obeys the Gaussian distribution with zero mean and 
unit variance.

\subsection{B. Numerical result for $d=1$}
\label{1d-result}

The numerical parameters used in the main text are $\Delta x=1, \cdots, 2.5$,
$\Delta t=0.0005$, $L=400$ and $L_y=100$ for the system with $\Gamma=0.1$,
$T=0.5$, $\kappa=1600$, $b_1=0.5$, $b_2=5.0$ and $\phi_0=-0.5$. Note that 
the correlation lengths of fluctuations in the bulk regions are $\xi_1=4.33$
and $\xi_2=27.3$. In this subsection, we show the numerical results in the
one-dimensional system to demonstrate that these numerical parameters
are appropriate.  The parameter values of the system are the same as the
two-dimensional case except for $L=800$. The numerical simulation method
for the one-dimensional system is basically the same as the two-dimensional
case.

\begin{figure}[H]
\centering
\includegraphics[width=80mm]{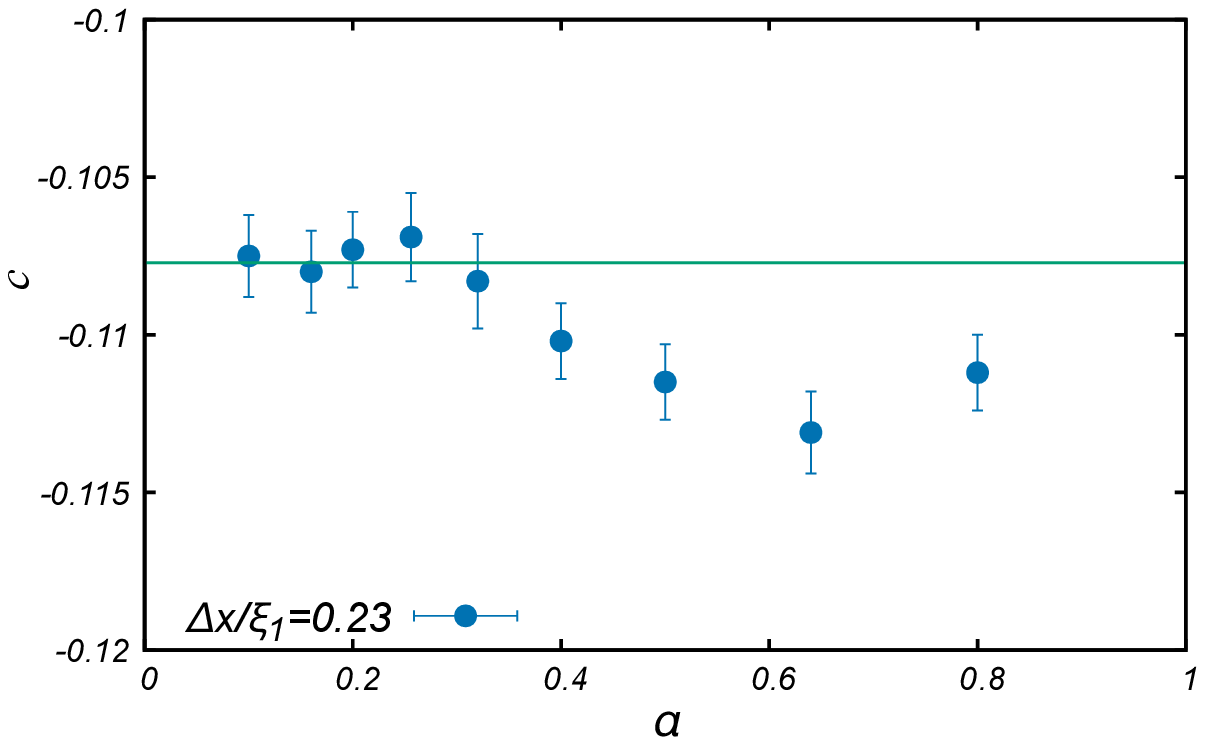}
\caption{
$\alpha$-dependence of the propagating velocity $c$ for the system
in one dimension. The blue circles represent numerical results.
The green line represents (8) in the main text.
Five hundred samples are used to estimate $c$ with error-bars.
}
\label{a_c_dx1}
\end{figure}

First, we study a time-interval dependence of the propagation velocity.
We define  a  dimensionless time interval 
\begin{equation}
\alpha \equiv 2 \kappa \Gamma \frac{\Delta t}{(\Delta x)^2}.
\end{equation}
The diffusion term causes the instability when $\alpha \ge 1$.
In Fig. \ref{a_c_dx1}, we show  the propagating velocity $c$
for various values of $\alpha$ with $\Delta x/\xi_1=0.23$ fixed.
We then find that the data for small $\alpha$ converges to the
formula of (8) in the main text. Since $\alpha \le 0.16$
for the numerical results presented in the main text,
we judge that our choice of the numerical parameters is
appropriate.

\begin{figure}[H]
\centering
\includegraphics[width=80mm]{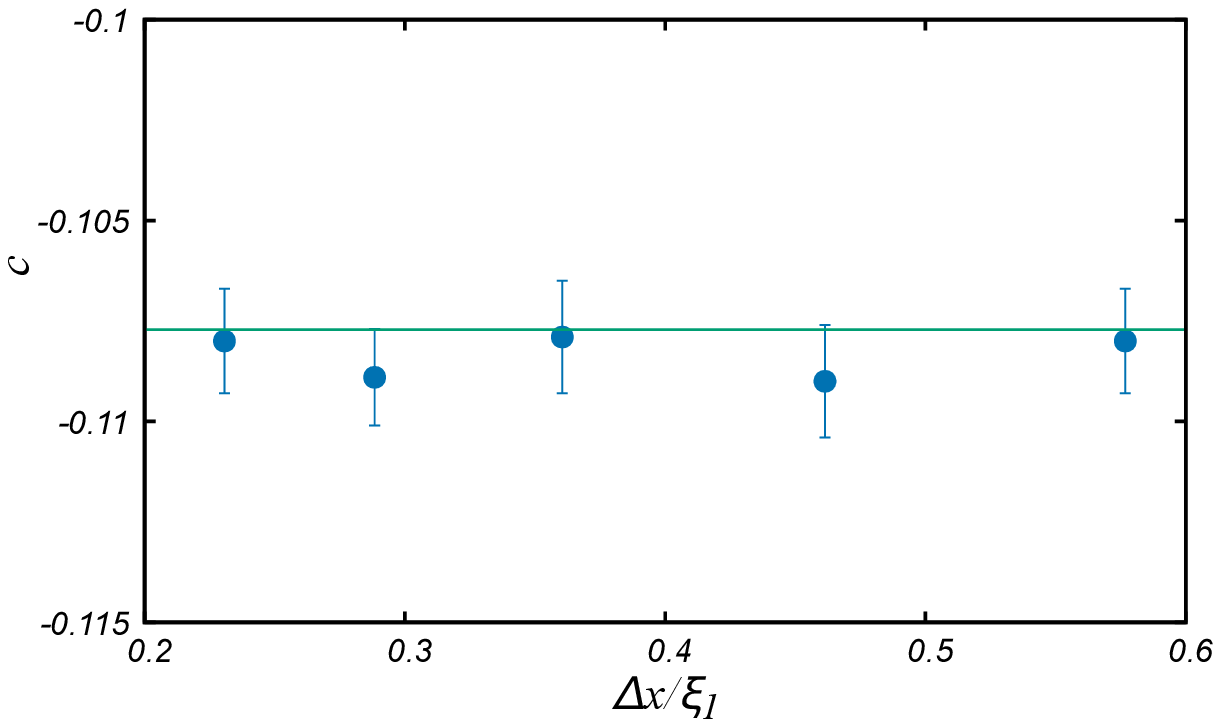}
\caption{
$\Delta x$-dependence of the propagating velocity $c$  for
the system in one dimensions. The blue circles represent numerical results.
The green line represents (8) in the main text.
Five hundred samples are used to estimate $c$ with error-bars. 
}
\label{1dim_dx_c}
\end{figure}

Next, we study a space-interval dependence of the propagation velocity.
In Fig. \ref{1dim_dx_c},  we show  the propagating velocity $c$
for various values of $\Delta x/\xi_1$ with $\Delta t =5.0 \times 10^{-4}$
fixed. Note that $\alpha$ takes smaller values than $0.16$ in the range
of $\Delta x/\xi_1$ we study. We find that the numerical result
is consistent with the theory. This gives another evidence that our
numerical simulations are reliable. 

\begin{figure}[H]
\centering
\includegraphics[width=80mm]{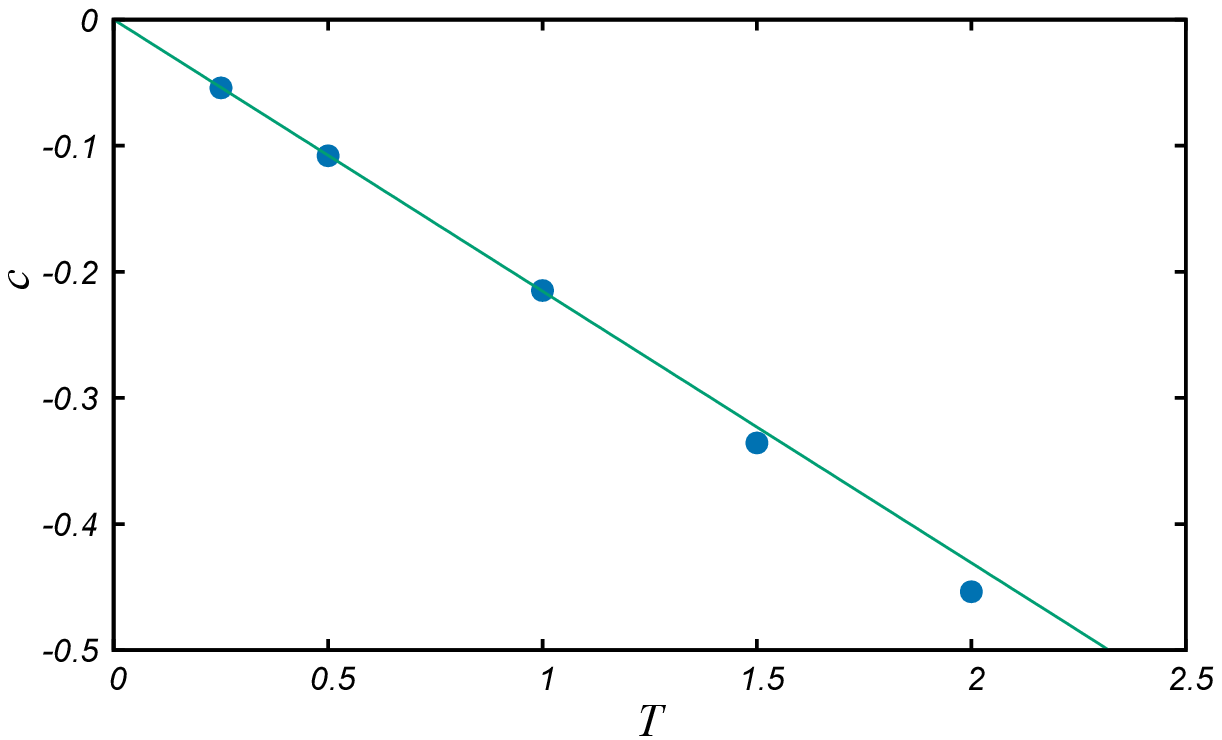}
\caption{
Temperature dependence of the propagating velocity $c$. 
The blue circles represent the numerical results
for the system with $\Delta x=1.0$ and  $\Delta t =5.0 \times 10^{-4}$. 
Five hundred samples are used to estimate $c$ with error-bars.
The green line represents the theoretical result (8) in the main text. }
\label{1dim_dx1_T_v}
\end{figure}

Finally, in Fig. \ref{1dim_dx1_T_v}, we show the temperature dependence
of $c$.  We observe that $c$ is proportional to $T$ when $T$ is less
than 1.0.

\subsection{C. Supplemental data for $d=2$}
\label{suport-d=2}
\begin{figure}[H]
\centering
\includegraphics[width=80mm]{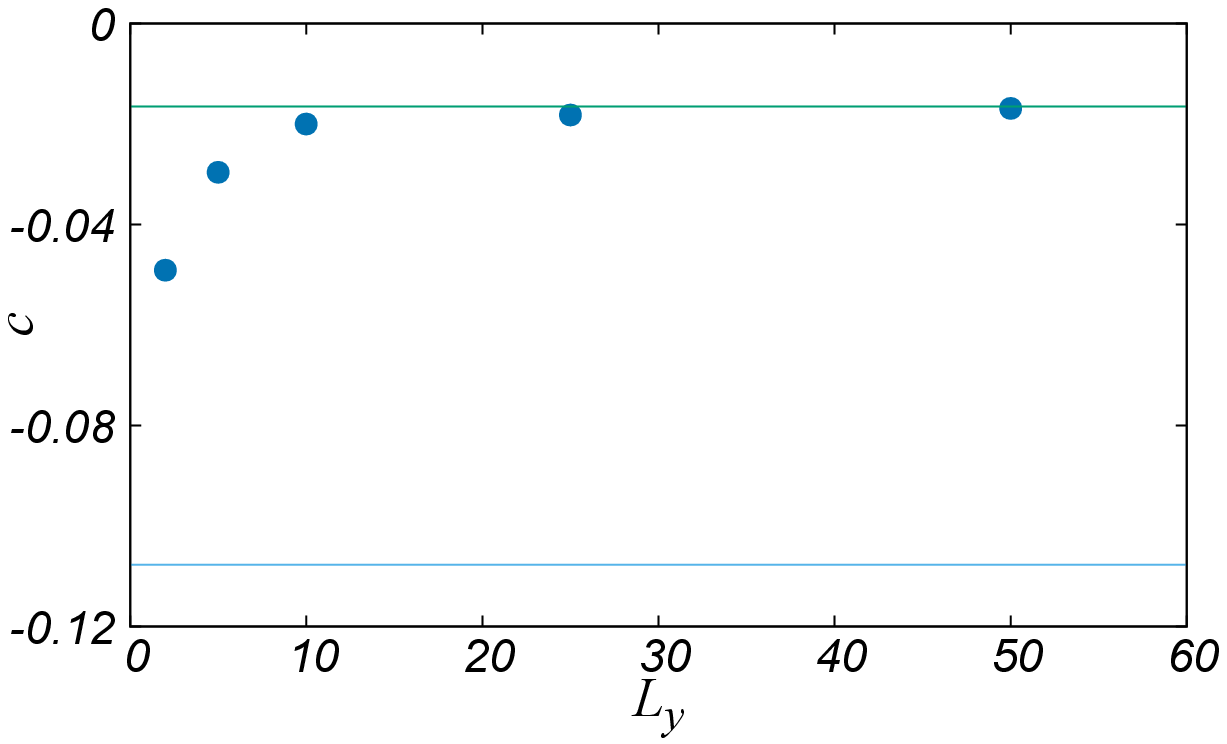}
\caption{
$L_y$-dependence of the propagating velocity $c$ for the
system in two dimensions. The blue circles represent numerical results.
The green line represents the formula (9) for the two-dimensional
system in the main text, while the light blue line represents the formula
for the one-dimensional system (8) in the main text. 
Fifty samples are used to estimate $c$ with error-bars.
The numerical parameter values are chosen as  $\Delta x=1.0$,
$\Delta t =5.0 \times 10^{-4}$ and $L = 400$.
}
\label{Ly_c}
\end{figure}

In this subsection, we investigate the parameter dependence
of the propagation velocity for the system with $d=2$.
We first study the $L_y$-dependence of $c$ to
check whether $L_y=100$ chosen for Fig. 4 is sufficiently large. 
In Fig. \ref{Ly_c}, we plot $c$ for various values of $L_y$.
We find that $c$ increases as $L_y$ increases and that $c$
converges to the value of the formula (9) of the
two-dimensional system in the main text. We thus judge that
$L_y=100$ is sufficiently large.

\begin{figure}[H]
\centering
\includegraphics[width=80mm]{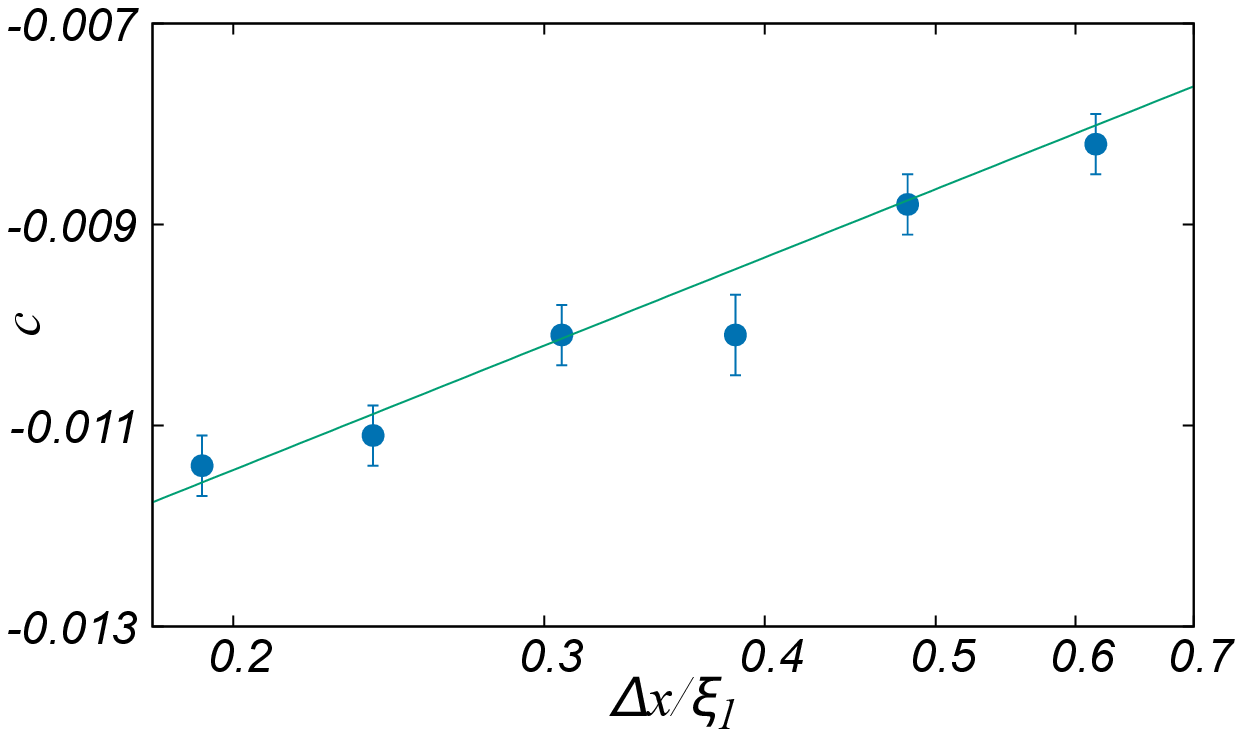}
\caption{
  Similar graph as Fig. 4 in the main text, where
  $\kappa=3600$ while the other parameter values are the same as
  those for Fig. 4. Eighty samples are used to estimate $c$
  with error-bars. }
\label{2dim_dx_c_k3600}
\end{figure}

\begin{figure}[H]
\centering
\includegraphics[width=80mm]{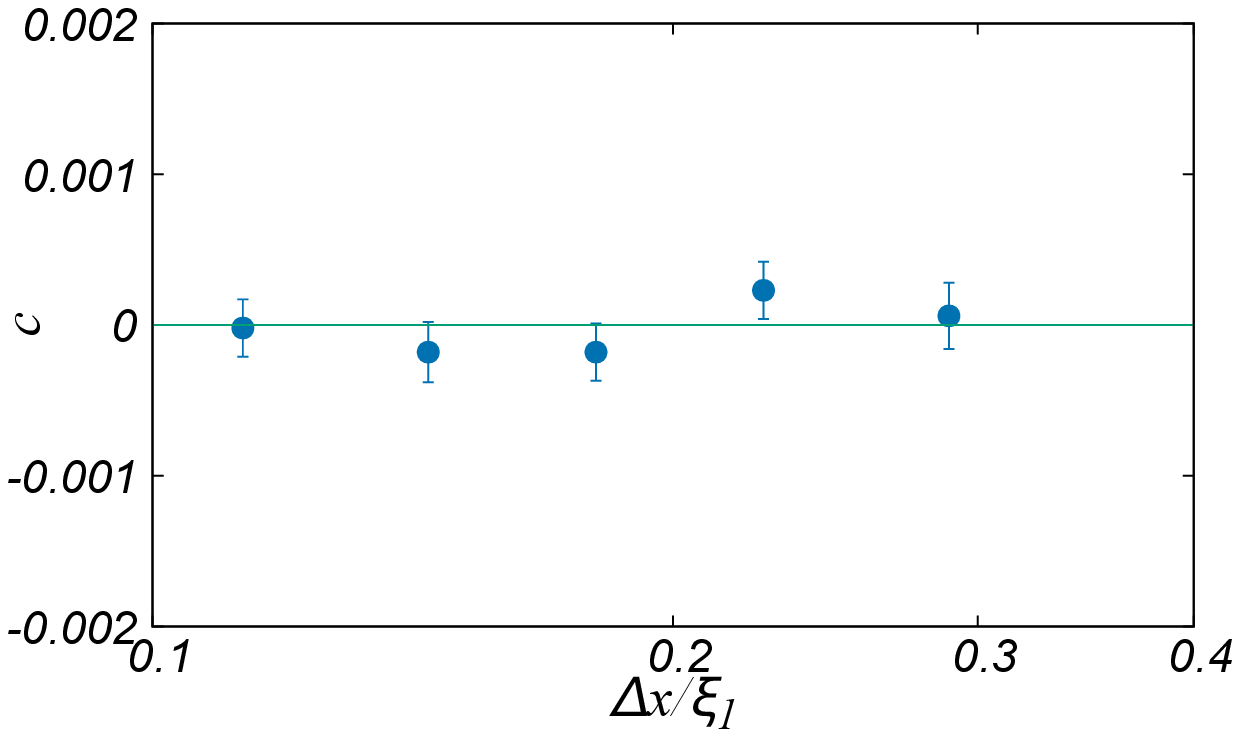}
\caption{
  Similar graph as Fig. 4 in the main text, where the symmetric
  potential is used while the other parameter values are the same as
  those for Fig. 4. One hundred samples are used to estimate $c$
  with error-bars.
}
\label{2dim_sym}
\end{figure}

Next, we study the propagation velocity for another value of
$\xi_1$ and $\xi_2$ to check the generality of the cut-off dependence.
The numerical parameters are the same as those in Fig. 4 in the main
text except for (i) $\kappa=3600$ or (ii) $b_1=b_2=0.5$, and $\phi_0=0$.
Note that the correlation lengths of fluctuations in the bulk regions
are $\xi_1=6.49$ and $\xi_2=40.95$ for the case (i) and 
$\xi_1=\xi_2=13.85$ for the case (ii). 
In Figs. \ref{2dim_dx_c_k3600} and \ref{2dim_sym},
we show  the propagating velocity $c$ for various values of
$\Delta x/\xi_1$ with $\Delta t =5.0 \times 10^{-4}$ fixed.
Note that $\alpha$ takes smaller values than $0.23$ for the case (i)
and $0.07$ for the case (ii). We find that the numerical results
are consistent with our theoretical calculation.

\end{document}